\documentclass[aps,prl,floatfix,superscriptaddress,twocolumn]{revtex4}
\usepackage{graphicx} \usepackage{epsfig}
\usepackage{amssymb}
\begin{document}
\title{Mesoscopic transport in ultrathin films of La$_{0.67}$Ca$_{0.33}$MnO$_3$}
\author{C. Beekman, J. Zaanen and J. Aarts}
\affiliation{Kamerlingh Onnes Laboratorium, Leiden University, The
Netherlands}
\email[E-mail:]{aarts@physics.leidenuniv.nl}
\begin{abstract}
We investigate the electrical transport in mesoscopic structures
of La$_{0.67}$Ca$_{0.33}$MnO$_3$ in the regime of the
metal-insulator transition by fabricating microbridges from
strained and unstrained thin films. We measure current-voltage
characteristics as function of temperature and in high magnetic
fields and with varying film thickness. For strained films, in
warming from the metallic to the insulating state, we find
non-linear effects in the steep part of the transition
characterized by a differential resistance with a strong peak
around zero applied current, and saturating at higher currents
after resistance drops up to 60 $\%$. We propose that this
nonlinear behavior is associated with melting of the insulating
state by injecting charge carriers, signalling the occurrence of
an intervening phase which involves the formation of short range
polaron correlations.
\end{abstract}
\pacs {} \maketitle
Doped manganites are strongly correlated electron systems which show a large variety
in physical properties. The material La$_{0.67}$Ca$_{0.33}$MnO$_3$(LCMO) for
instance shows a transition from a paramagnetic insulator to a ferromagnetic metal
around $T_{MI}$ = 250 K and the well known colossal magnetoresistance (CMR) effect.
The physics is mainly determined by competing interactions: trapping of the
electrons in Jahn-Teller (JT) distortions (polarons) and the itinerancy of the
electrons in the Double-Exchange mechanism \cite{zener4} when spins become
polarized. This competition signals that only small free energy differences exist
between a variety of different possible phases of the system.  As a result the phase
of the material can be tuned easily by various external perturbations, such as
magnetic and electric fields, strain and disorder. For instance, the application of
strain amplifies the JT splitting of the e$_g$ levels and makes the distortions more
static in nature. This has an inhibiting effect on band formation, which leads to
reduction of T$_{MI}$ in manganite thin films compared to the bulk value
\cite{freisem4, dorrrev4}. Furthermore, the susceptibility of the M-I transition to
disorder (doping disorder, oxygen nonstoichiometry, defects from strain relaxation,
twinning, and grain boundaries) can lead to the coexistence of the metallic and
insulating phases
on a variety of length scales \cite{uehara2,lynn5}. \\
Electrical transport in these correlated electron systems is therefore a complex
phenomenon, which has however hardly been probed on small length scales. Here we
address such transport in the mesoscopic regime, by investigating microbridges made
in LCMO thin films, grown in a strained or unstrained state on SrTiO$_3$ (STO) or
NdGaO$_3$ (NGO) substrates. In the strained films we do find strongly nonlinear
transport behavior in the CMR regime near the metal-insulator transition,
characterized by increased conductance at higher current. We attribute these effects
to the intrinsic physics of the insulating state in terms of the formation of an
intervening glassy polaron state when going from the correlated metal to the
polaronic insulator.

\begin{figure}
\includegraphics[width=8.5cm]{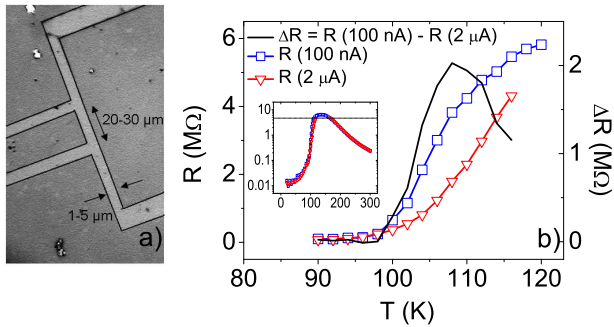}
\caption{(color on-line) a) electron microscopy picture of a
typical microbridge. The width of the bridge is 5~$\mu$m. b)
$R(T)$ between $T$ = 90 and 120 K at two different current
densities (\emph{J} = 2x10$^6$ A/m$^2$ (squares) and \emph{J} =
4x10$^7$ A/m$^2$ (triangles)) for the 10 nm thick microbridge on
STO (left axis). (right axis and solid line) the difference
between the two curves, $\Delta$R = R(100nA)-R(2$\mu$A). Inset:
$R(T)$ between $T$ = 20 and 300 K for both current densities on a
logarithmic scale. The dashed line indicates the measurement limit
for the high current density measurement. The absence of
resistance values above this limit is due to saturation of the
nanovoltmeter.} \label{fig1}
\end{figure}

We have grown LCMO thin films with varying thicknesses between 7 - 20 nm on STO and
NGO substrates, by DC-sputtering in an oxygen atmosphere of 3 mbar and at a growth
temperature of 840$^{\circ}$C. The STO induces a tensile strain in the LCMO film due
to the lattice mismatch of -0.6 $\%$,  the NGO is (almost) lattice matched. After
growth, x-ray measurements were performed to check the thickness of the films.
Furthermore, Electron Energy Loss Spectroscopy (EELS) was used to characterize the
elemental composition and Mn valence state of the samples \cite{beekman2}. We find
that our films have compositions close to that of the sputtering target and correct
oxygen stoichiometry. The films were patterned using electron-beam lithography and
conventional Ar-etching followed by an oxygen plasma etch to restore the insulating
properties of the substrate \cite{beekman}. The mesoscopic LCMO structure has a
four-point configuration with dimensions of 5 $\mu$m (width) by 20 - 30 $\mu$m
(length between voltage contacts), as shown in Fig.\ref{fig1}a. Macroscopic Au/MoGe
contacts were fabricated in a second e-beam step. We measured $I$-$V$ curves as
function of temperature and in high magnetic fields using a Physical Properties
Measurement System (Quantum Design) for temperature control ($T$ = 20 - 300~K) and
for magnetic field control ($H_a$ = 0 - 9~T), but with external current sources and
voltmeters.

$I$-$V$ characteristics, measured by varying the current, were obtained across the
temperature range of 20 - 300~K in order to obtain the differential resistance. All
microbridges were also measured in a magnetic field of 5~T to investigate the
magnetoresistance (MR) effect. Our main findings are shown in Fig.\ref{fig1}b. It
shows the resistance vs. temperature ($R(T)$) of a microbridge on STO, measured at
currents of 100~nA (current density \emph{J} = 2x10$^6$ A/m$^2$) and 2~$\mu$A
(\emph{J} = 4x10$^7$ A/m$^2$). As can be seen in the inset, where $R(T)$ is plotted
logarithmically, both sets of data show a clear M-I transition typical for strained
LCMO thin films with $T_{MI}$ = 130 K (peak temperature) and a resistance drop of
three orders of magnitude.
The main panel of Fig.\ref{fig1}b shows $R(T)$ on a linear scale. The higher current
density results in a reduction of resistance just below T$_{MI}$, which produces a
shift in the upper part of the transition of 5 to 10~K. Also shown is the difference
between the two curves, which peaks between 105~K and 110~K. If we define a
current-induced resistance called Electroresistance (ER($\%$) = $\frac{R_{high} -
R_{low}}{R_{low}}\times 100\%$), then we observe a maximum ER effect up to 60 $\%$
for the 10 nm microbridge.

\begin{figure}
\includegraphics[width=8cm]{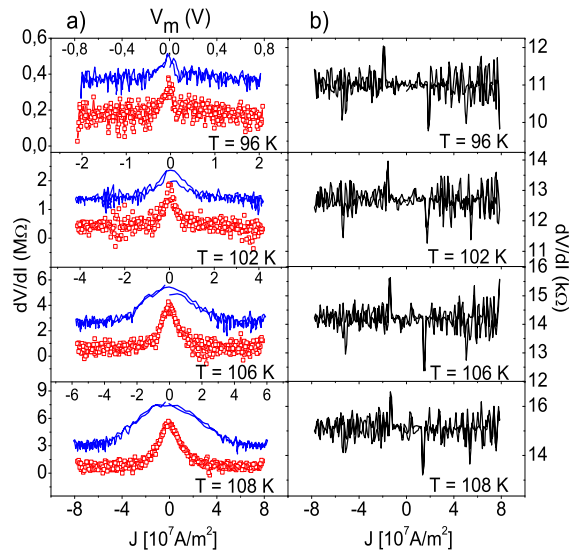}
\caption{(color on-line) The numerical derivatives of the $I$-$V$
curves for the 10 nm thick microbridge on STO at four different
temperatures of which the R(T) behavior is shown in Fig.
\ref{fig1}. a) Measured in 0 T. Bottom axis shows current density,
\emph{J} (symbols) and top axis indicates the measured potential
difference (V$_m$) between the voltage contacts (solid line). b)
Measured in 5 T as function of current density.}\label{fig2}
\end{figure}

\begin{figure}
\includegraphics[width=7cm]{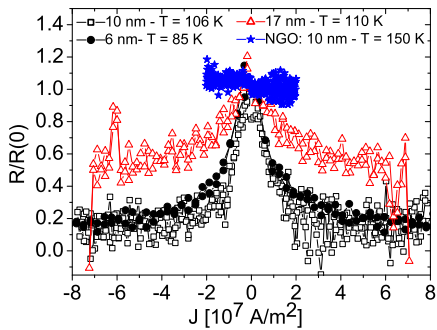}
\caption{(color on-line) Numerical derivatives dV/dI of the
$I$-$V$ curves for three different bridges on STO. The curves are
normalized w.r.t. the zero bias resistivity. Film thicknesses are
indicated as well as the temperature at which the $I$-$V$ was
measured. For all films the curves are shown for the temperature
were the largest reduction as function of applied current
occurred. The zero bias resistances are, d = 17 nm: R(0) = 0.3
M$\Omega$, d = 10 nm: R(0) = 4.0 M$\Omega$ and d = 6 nm: R (0) =
4.3 M$\Omega$. Also shown is the $\frac{dV}{dI}$ curve at T = 150
K of the 10 nm LCMO film on NGO which does not show nonlinear
behavior.}\label{fig3}
\end{figure}

The measured $I$-$V$ curves are linear for microbridges on NGO and linear for most
temperatures in the case of STO. However, for microbridges on STO we find nonlinear
behavior in the steep part of the transition which corresponds to the behavior shown
in Fig.\ref{fig1}. Typical nonlinear behavior is shown in Fig.\ref{fig2}.
The left panel shows the differential resistance for the 10 nm microbridge at four
different temperatures in 0~T as function of applie
d current density (bottom axis)
and also as function of the measured potential difference (V$_m$) between the
voltage contacts (top axis). At low temperature all $I$-$V$ curves are linear up to
$J$ = 8x10$^{7}$ A/m$^2$ (I = 4 $\mu$A). Upon warming into the transition the
nonlinear behavior starts to occur just below T = 96~K and appears to continue until
T$_{MI}$. For all microbridges the differential resistance is largest at zero bias
and drops with increasing applied current density. The full width of the peak
appears to increase by more than an order of magnitude from about 2x10$^{6}$ A/m$^2$
(0.2 V) at T = 96 K up to 4x10$^{7}$ A/m$^2$ (8 V) at T = 110 K. However, due to the voltage
limit of the nanovoltmeter, it is difficult to observe any nonlinear behavior
between 110 K - 170 K. Above this temperature range all $I$-$V$ curves are linear.
From the right panel it is clear that application of a high magnetic field leads to
a reduction in the differential resistance and to complete disappearance of the
nonlinearities in the $I$-$V$ curves; they are linear across the entire temperature
range. To compare the differential resistance behavior of different microbridges we
show ($\frac{dV}{dI}$, Fig.\ref{fig3}), normalized to the value at zero bias, as function of current
density $J$ for three microbridges on STO with different thicknesses.  The film thickness is indicated for each curve as well as the
temperature at which the $I$-$V$ curve was measured.
The variation in ($\frac{dV}{dI}$) with $J$ becomes less strong
when the microbridge thickness is increased. Also shown in
Fig.\ref{fig3} are similar data of a 10 nm LCMO film grown on an
NGO substrate (T$_{MI}$ = 165 K) measured at 150 K, again in the
steep part of the transition. Although measured in a somewhat
smaller current density range we find no nonlinear behavior in
films on NGO for thicknesses down to 10 nm.

\begin{figure}
\includegraphics[width=10cm]{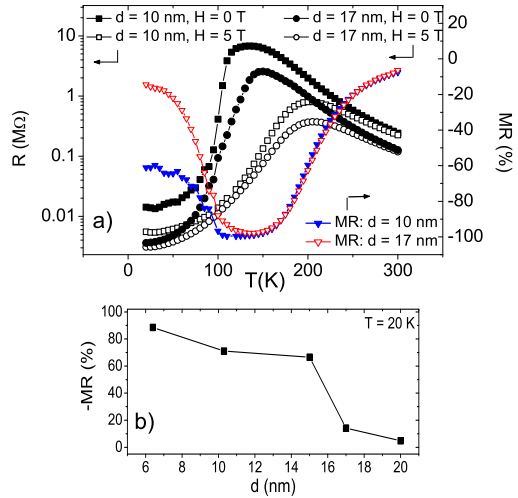}
\caption{(color on-line) a) R(T) behavior in 0 T and 5 T of a 17
nm thick and 10 nm thick microbridge. The calculated
magnetoresistance as function of temperature is also shown. b) The
calculated magnetoresistance (MR ($\%$) = $\frac{R_{5T} -
R_{0T}}{R_{0T}}\times 100\%$) as function of film thickness at T =
20 K.} \label{lowTMR}
\end{figure}

Besides the well known CMR effect in the transition, we also observe a strong MR
effect at low temperatures both for microbridges on STO and on NGO. As becomes clear
from Fig.\ref{lowTMR}a this effect too depends on the microbridge thickness. For the
17 nm thick bridge the $R(T)$ curves at 0~T and 5~T almost overlap, but for the 10
nm thick bridge a high magnetic field induces a significant reduction in resistance
at low temperature. The magnitude of the low temperature MR increases with reducing
film thickness reaching about an order of magnitude for the 6.4 nm thick film, see
Fig.\ref{lowTMR}b. Furthermore, also the (unstrained) 10 nm thick film on NGO shows
a MR effect, of 50~$\%$ at low temperature. Apparently, even for the unstrained LCMO
microbridge an applied magnetic field can result in an increased metallicity at low
temperature. On the other hand, we stress that in this regime of enhanced MR the
$I$-$V$ characteristics are simply ohmic. \\

Behavior as found here under the controlled circumstances of substrate, sample
thickness, and bridge width, has not been reported before. Nonlinear behavior was
observed recently in thick unstructured and strongly oxygen deficient samples
\cite{liu5}, but temperature regime, magnetization and field dependence were all
different, and it is difficult to connect that work to our observations, in
particular since the EELS characterization showed that oxygen deficiency is not an
issue in our LCMO films. A concern can be that the observations are an artefact
caused by Joule heating in the microbridge. The peak resistivity is around $\rho$
$\sim$ 10$^5$-10$^6$ $\mu\Omega$ cm which we can use to estimate the effect of Joule
heating in the measured current density range. The power inserted into the bridge is
of the order of $\mu$W; the estimated Joule heating would be in the mK range which
is clearly negligible. Furthermore, heating would lead to different nonlinear
behavior, namely increasing resistance with increasing current, and can be ruled out
as a possible cause for the observed nonlinearities. Another concern could be the
influence of the structural phase transition, tetragonal to cubic, which occurs in
the STO substrate at $T$ = 105~K . Since our LCMO films are epitaxial this could
influence microbridge properties. However, we have shown that the observed nonlinear
behavior can occur at different temperatures, both above and below T = 105 K. This
indicates that the nonlinear behavior is not an STO-induced effect but intrinsic to
the material LCMO. \looseness-1

\begin{figure}
\includegraphics[width=9cm]{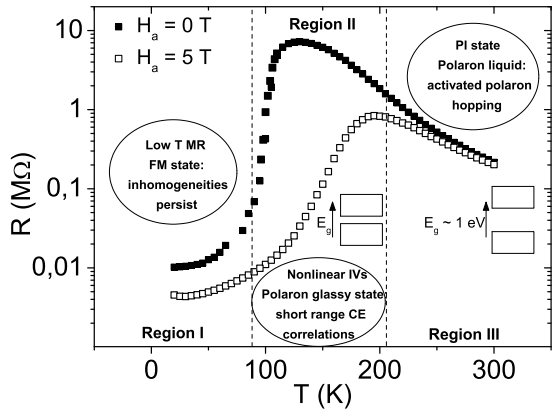}
\caption{Proposed scheme for the metal-insulator transition in LCMO (R(T) for the 10
nm microbridge at $J$ = 2x10$^6$ A/m$^2$). Region I: at low temperature the material
is in a FM state but nanoscale inhomogeneities persist. Region II: in the transition
e$_g$ electrons are localized by Jahn-Teller splitting resulting in the formation of
a glassy state of correlated (CE-type, charge ordered) polarons. Region III: at a
certain temperature above the transition the polaron correlations break down and the
system becomes a polaron liquid with only single dynamic polarons. In the graph the
JT-splitting is indicated with E$_g$ $\sim$ 1 eV in the polaron liquid
state.}\label{dis5}
\end{figure}

We conclude from this that the non-linear behavior has to be the
fingerprint of an organized phenomenon intrinsic to the electron
matter formed in the manganite.  Next we discuss the different
regimes we believe are present, c.q. the different states of the
microbridge as it is warmed through the metal-insulator
transition. In Fig.\ref{dis5} we provide the R(T) data of the 10
nm microbridge again, in order to help in identifying the various
regimes. At low temperature (region I), the strong MR effects show
that a high magnetic field can still assist in increasing the
metallicity of the microbridge. We believe this to derive from
static inhomogeneities which lead to a relatively high residual
resistance in the thinnest films and locally frustrate the DE-type
metallic state which then forms easier in a (high) magnetic field.
The effect is rather similar to the MR reported in
Ref.\cite{eblen05} in thin films of La$_{0.8}$Ca$_{0.2}$MnO$_3$
which also exists down to the lowest temperatures.\\
Upon warming into the transition the conduction electrons become
more localized, with the Jahn-Teller splitting of the
e$_g$-levels, assisted by the strain in the film, leading to
polaron formation. In the steep part of the transition we start to
observe strongly nonlinear behavior, and increasing conductance
with current. The nonlinearities are fully reversible, indicating
that the process which enhances the conductance is not a
first-order transition. A scenario in which current transforms a
possible antiferromagnetic insulator to a ferromagnetic metal,
e.g. through spin torque processes, is not likely, since the
closeness of the M-I transition in LCMO to a first-order
transition \cite{kim5} would probably render this process
hysteretic. The scenario we propose instead is that of
current-induced melting of an intervening phase which can sustain
a voltage difference, while its electrical properties are
extraordinarily sensitive to the injection of charge carriers as
occurs in the high current state. Such a phase was actually
demonstrated to exist, and is called the polaron glass phase
\cite{lynn5}. We can understand the origin of this phase by
remembering that at higher doping of x $\geq$ 0.5, the material is
antiferromagnetic and charge and orbital ordered. At lower doping
x = 0.33 (our material), this ordering is frustrated, but polaron
correlations can still occur. In ref.\cite{lynn5} it was shown
through neutron scattering experiments that a correlated polaron
(glass) phase is formed in LCMO single crystals. The nanoscale
structural correlations occur just above T$_{MI}$. The development
of these static (charge ordered, CE-type) polaron structures trap
electrons and drive the system into the insulating state. In our
case the correlated regions already start to occur below T$_{MI}$
and become more abundant when the bridge is warmed through the
transition, a process facilitated by the strain, which causes more
disorder on nanoscales as well as larger Jahn-Teller distortions
and smaller band-widths. The resulting glass phase fully closes
the bridge off. The injection of carriers into this nascent state
by applying a chemical potential difference works against the
formation of the polaron correlations and drives the system to a
different, more metallic, equilibrium. We note that the large CMR
effect in the transition is in itself related to the occurrence of
the correlated polaron phase, as reported both experimentally
\cite{kiryukhin5,tendeloo5} and theoretically \cite{sen5}, while a
similar scenario was suggested for the bilayer compound
La$_{1.2}$Sr$_{1.8}$Mn$_2$O$_7$ \cite{mannella5}. Much weaker CMR
effects are observed in systems with only single dynamic polarons
(large bandwidth materials such as La$_{0.67}$Sr$_{0.33}$MnO$_3$).
Warming into region III, the polaron correlations break down
(polaron liquid) and conduction is governed by thermally activated
(single) polaron hopping.

In summary, we find that, upon reducing the size of a strained LCMO film grown on an
STO substrate, novel behavior in the transport properties occurs, notably non-linear
current-voltage characteristics. This is not found in wider bridges or when strain
is absent (films on NGO). As a possible explanation we use the concept of a phase of
glassy polarons which is formed in the M-I transition, assisted by the strain, and
which is very sensitive to the injection of charge carriers, leading to
current-induced melting of the newly forming insulating state.

We thank I. Komissarov for discussions, and M. Porcu and H. Zandbergen for
performing the EELS measurements. This work was part of the research program of the
Stichting voor Fundamenteel Onderzoek der Materie (FOM), which is financially
supported by NWO.

\end{document}